\documentclass[
 reprint,
 superscriptaddress,
 amsmath,amssymb,
 aip,
 jcp
]{revtex4-2}

\usepackage{graphicx}
\usepackage{dcolumn}
\usepackage{bm}
\usepackage{gensymb}
\usepackage{booktabs}
\usepackage{hyperref}
\usepackage{amsmath}
\newcommand{\angstrom}{\,\textup{\AA}}
\usepackage{float}

\usepackage{xcolor}
\usepackage{soul}

\renewcommand{\vec}{\textbf}

\begin{document}

\title{Exploring the configurational space of amorphous graphene with machine-learned atomic energies}

\author{Zakariya El-Machachi}
\affiliation{Department of Chemistry, Inorganic Chemistry Laboratory,
University of Oxford, Oxford OX1 3QR, UK}
\author{Mark Wilson}
\affiliation{Department of Chemistry, Physical and Theoretical Chemistry Laboratory,
University of Oxford, Oxford OX1 3QZ, UK}
\author{Volker L. Deringer}
\email{volker.deringer@chem.ox.ac.uk}
\affiliation{Department of Chemistry, Inorganic Chemistry Laboratory,
University of Oxford, Oxford OX1 3QR, UK}

\date{\today}

\begin{abstract}
Two-dimensionally extended amorphous carbon (``amorphous graphene'') is a prototype system for disorder in 2D, showing a rich and complex configurational space that is yet to be fully understood. 
Here we explore the nature of amorphous graphene with an atomistic machine-learning (ML) model. 
We create structural models by introducing defects into ordered graphene through Monte-Carlo bond switching, defining acceptance criteria using the machine-learned {\em local}, atomic energies associated with a defect, as well as the nearest-neighbor (NN) environments. 
We find that physically meaningful structural models arise from ML atomic energies in this way, ranging from continuous random networks to paracrystalline structures. 
Our results show that ML atomic energies can be used to guide Monte-Carlo structural searches in principle, and that their predictions of local stability can be linked to short- and medium-range order in amorphous graphene. 
We expect that the former point will be relevant more generally to the study of amorphous materials, and that the latter has wider implications for the interpretation of ML potential models.
\end{abstract}

\maketitle

\section*{\label{sec:Intro}Introduction}

The study of the amorphous state has long been of fundamental research interest.\cite{Wright2014} It is also increasingly important to understand structure--property correlations in amorphous materials, owing to ubiquitous applications in solar cells,\cite{SHI2015} transparent electronic devices,\cite{Zhao2012} or phase-change memories.\cite{Le_Gallo_2020}
Whilst bulk amorphous phases are challenging to study structurally, two-dimensional (2D) amorphous materials can be directly visualized by atomistic imaging techniques such as high-resolution transmission electron microscopy (HRTEM).\cite{Kotakoski2011, Huang2012, Huang2013, yang_field-effect_2015, Joo2017, yang_synthesis_2020, boron_nitride} And just like graphene is the prototypical ordered 2D material, there is ongoing research interest in its disordered analogue(s). Indeed, Toh et al.\ recently synthesized a centimeter-scale sample of free-standing monolayer amorphous carbon\cite{toh_synthesis_2020} and characterized the structure based on the interpretation of HRTEM images.  

The amorphous forms of carbon have been widely studied using computer simulations. Most commonly, these studies have been carried out in the framework of molecular dynamics (MD), from early work on melt-quenching \cite{Tersoff_1988, Galli1989, Drabold1994, Marks1996, McCulloch2000, Wilson2012, Wilson2013} to direct simulations of thin-film growth by ion deposition. \cite{Kaukonen1992, Marks2005, Caro2018, Caro2020} Very recently, the properties of ``amorphous graphite'', as a 3D extended arrangement of individual amorphous graphene sheets, were investigated with density-functional theory (DFT) based MD and electronic-structure computations.\cite{amorphgraphite} Earlier simulation studies had emphasized the connection between low-density amorphous carbon and the idealized case of 2D amorphous graphene (``aG'' in the following). \cite{Bhattarai2018, Bhattarai2018a}

Graphene itself, as a 2D system, is rather well-confined and allows for Monte Carlo (MC) simulations to model topological defects. The Wooten--Winer--Weaire (WWW) algorithm, initially proposed for silicon,\cite{WWW} remains a simple yet robust approach to generating continuous random network (CRN) models. In this, disorder is gradually introduced through local bond transpositions and structural relaxation. To simulate aG, bond transpositions are introduced as Stone--Wales (SW) defects,\cite{SW} with moves being accepted or rejected based on a suitable Metropolis criterion, typically the total-energy difference between the new and old configuration. SW defects are created by a formal in-plane $90\degree$ rotation of two atoms around the mid-point of the bond\cite{SW} and are the foundational example of (intrinsic) topological disorder in 2D carbon.\cite{Meyer2008, SW_DFT, Banhart2011, Thiemann2021} This approach has long been used to study aG. \cite{Kapko2010} Various implementations of the WWW algorithm exist: D'Ambrosio et al.\ showed that the majority of bond transpositions are rejected during annealing, and that an early decision scheme can enhance computation speed by rejecting unfavorable transpositions;\cite{dambrosio_efficient_2021} Ormrod Morley et al.\ constructed Metropolis criteria from topological metrics such as ring distributions (that is, energy changes were not considered in that case).\cite{ormrod_morley_controlling_2018}

Machine learning (ML) based interatomic potential models are increasingly being used to accelerate materials simulations. \cite{Behler2017, Deringer2019, Friederich2021} ML potentials are typically ``trained'' with DFT data and can achieve similar accuracy for a small fraction of the cost. One key assumption in many of these methods is that the total energy can be separated into sums of machine-learned atomic energies. \cite{Behler2007,Bartok2010}
Whilst being an approximation in the first place, it was argued that these atomic energies may in fact be amenable to interpretation: 
a connection between ML atomic energies and local chemical structure was made for partly occupied crystallographic sites in $\beta$-rhombohedral boron,\cite{Boron_local} and for atomic environments with different coordination numbers in amorphous silicon.\cite{Si_local} 
ML models for other properties of local atomic environments have recently been investigated, ranging from the local electronic density of states \cite{BenMahmoud2020, deringer_origins_2021} to local distortion factors in grain boundaries.\cite{Atomic_GB_Local} The nature of these local ML properties (including atomic energies), and their usefulness in predicting physical properties, remains an interesting research question. (See, e.g., Ref.\  \citenum{Behler2019} for a discussion of ML atomic energies in a chemically complex system.)

In the present work, we explore the configurational space of amorphous graphene based on an ML potential model that gives access to total and local energies. We use an MC bond-switching algorithm where ML atomic energies are used in the acceptance criterion and show that doing so leads to physically sound structural models. Depending on the details of the algorithm, we obtain CRN-like or paracrystalline structures. Our work shows that ML atomic energies can be used in different ways to ``drive'' MC simulations based on local and nearest-neighbor (NN) energy contributions, with implications for research on amorphous graphene and likely on other disordered structures and materials. 

\section*{\label{sec:Methods}Methods}

\subsection*{\label{sec:PE_models}Potential-energy models}
A common \textit{ansatz} in developing potential-energy models is that the total energy, $E$, can be separated into a sum of atomic contributions:
\begin{equation} \label{total_e}
    E=\sum_{i}^{\text{atoms}}{\varepsilon_{i}},
\end{equation}
where $\varepsilon_{i}=\varepsilon(\{\textbf{r}_{ij}\})$, with $i,j$ being atomic indices, and $r_{ij} \le r_{\rm cut}$. This \textit{ansatz} can be applied to many systems as short- and medium-range interactions predominantly determine the total energy. However, the question of how to formulate $\varepsilon(\{\textbf{r}_{ij}\})$ is not trivial: in carbon, there is a vast configurational space with a subtle interplay between structure and energetics, such as in dihedral and torsional forces, weak interlayer interactions, and so on. Moreover, locality can depend on the structure: numerical experiments showed a large difference in the locality of atomic forces for diamond versus graphite. \cite{GAP17}

Empirical potentials are typically parameterized for a specific composition and phase of a material. For example, the original Tersoff potential for carbon\cite{Tersoff_1988} was parameterized by fitting parameters of repulsive and attractive pair potentials to cohesive energies of carbon polymorphs along with the lattice parameter and bulk modulus of diamond. 
This potential\cite{Tersoff_1988} used a bond-order approach, where bond strengths are modified according to the number of neighbors. The original reactive empirical bond-order (REBO) potential was an update of the Tersoff potential incorporating hydrogen,\cite{brenner_empirical_1990} with REBO-II adding further improvements.\cite{brenner_second-generation_2002} A long-range Lennard-Jones term was added to REBO-II, creating the adaptive intermolecular reactive bond order (AIREBO) potential.\cite{stuart_reactive_2000} The long-range bond order potential (LCBOP)\cite{los_intrinsic_2003} is similar to AIREBO, having a long-range term in conjunction with a bond-order description---albeit here it is built in from the beginning. The environment-dependent interaction potential (EDIP) for carbon\cite{marks_generalizing_2000,Marks2002} was initially developed from an earlier silicon EDIP model.\cite{justo_interatomic_1998,bazant_environment-dependent_1997} EDIP has been shown to be successful in describing various properties of amorphous carbon, including the graphitization at low and high densities. \cite{de_tomas_graphitization_2016, DeTomas2017, Shiell2018}

\subsection*{\label{sec:GAP}Gaussian approximation potential (GAP)}

The Gaussian approximation potential (GAP) framework \cite{Bartok2010} is used to ``machine-learn'' interatomic potential models from quantum-mechanical data, often based on DFT. Unlike empirical potentials which are constructed based on physical knowledge, GAP makes a non-parametric fit. This means that the model can adjust to complex input data---however, it also means that the selection and quantity of reference (``training'') data is critically important.\cite{chemrev_GPR} 

In brief, the local energy for a GAP model is
\begin{equation}
 \varepsilon(\vec{q})=\sum_{t=1}^{N_{t}}\alpha_{t}K(\vec{q},\vec{q}_{t}),
\end{equation}
where the sum runs over $N_{t}$ training configurations represented by the local-environment descriptor $\textbf{q}_{t}$, with a corresponding weighting coefficient $\alpha_{t}$ attributed during fitting. $K$ is a covariance kernel which measures the similarity between the input and training configurations, represented by $\textbf{q}$ and $\textbf{q}_{t}$, respectively. A commonly used approach for the latter task is the Smooth Overlap of Atomic Positions (SOAP) descriptor and kernel. \cite{SOAP2013} 

For the present work we use the amorphous carbon potential, GAP-17.\cite{GAP17} This model has been shown to predict energies within tens of meV per atom compared to DFT, as well as providing a good description of structural and elastic properties of amorphous carbon.\cite{GAP17}
 
\subsection*{\label{sec:Init}Initialization}

We generate a pristine, 200-atom layer of cG using the experimental bond length of $1.42 \angstrom$ (Refs.\ \citenum{Structure_of_graphene_and_its_disorders} and \citenum{Experimental_Review_of_Graphene}). A spacing of $20 \angstrom$ between layers ensures that there is no interaction between periodic replicas. Where noted, NN-energy-based runs start from structural models that have been ``thermalized'', i.e., disordered, using local energies at $\beta=2.0$ eV$^{-1}$ (see below). The cell parameters are allowed to relax prior to the first bond switch, and then kept fixed for the duration of the MC simulation. The structures are kept planar, simplifying the present study to the idealized 2D case, and noting that puckering can lower the energy further. \cite{Li2011}

\begin{figure}
    \centering
    \includegraphics[width=\linewidth]{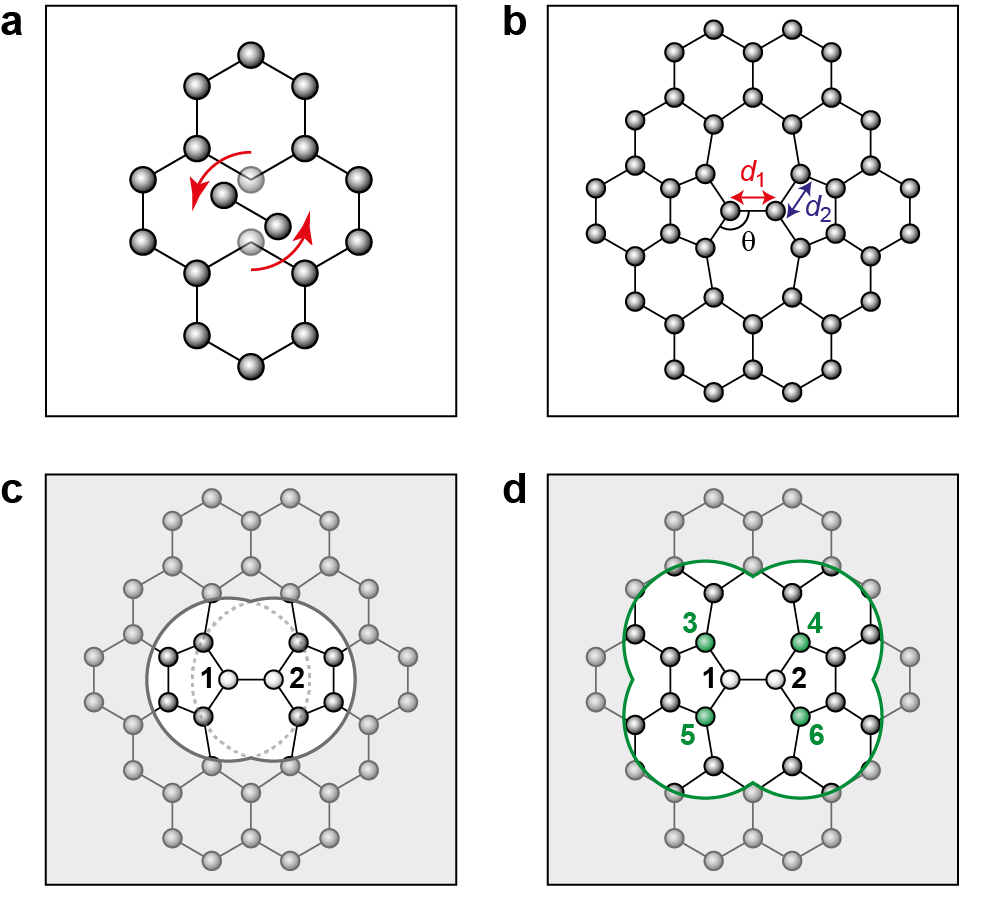}
    \caption{Stone--Wales defect in graphene and definition of nearest-neighbor (NN) atoms.
    (a) Schematic of an in-plane single-bond rotation in graphene, leading to an SW defect. (b) A bond rotation by 90$\degree$ creates two 5-membered and two 7-membered rings. Definitions are given for the defect bond length, $d_{1}$, the defect--NN bond length, $d_{2}$, and the defect bond angle, $\theta$ (cf.\ Table I). (c) Definition of local energy for the Metropolis criterion. The gray atoms labeled 1 and 2 are the SW defect pair, with shading indicating the overlap of the two cutoff spheres up to which atoms contribute to the local energy (this value is 3.7 \AA{} for the GAP-17 model, larger than sketched here). (d) The green atoms (3--6) are the topological NNs of the defect pair. Green lines indicate the overlap of the cutoff spheres for each NN atom.}
    \label{fig:cartoon}
\end{figure}

\subsection*{\label{sec:MC_Protocol}Monte-Carlo protocol}

A kinetic MC algorithm is used to generate aG structural models. The simulations begin either from cG or from a thermalized structure. At each step, a random atom in the $xy$ plane is chosen along with a random neighbor, determined using a cutoff of $1.85 \angstrom$. The atoms undergo an SW rotation ($90\degree$ about the bond center) and the new structure is then relaxed using the conjugate-gradient algorithm. For initial testing of structural relaxations, energy evaluations, and MC runs, we used the Atomic Simulation Environment (ASE) \cite{Hjorth_Larsen_2017} interfaced to \texttt{quippy}\cite{Csanyi2007-py,Kermode2020-wu} (\url{https://github.com/libAtoms/QUIP}). The production MC runs shown in this work used LAMMPS.\cite{LAMMPS} The force tolerance for structural relaxations was 1 meV$\angstrom^{-1}$ with a maximum of 150 relaxation steps. A topological constraint is imposed where the newly relaxed structure must be 3-coordinate, otherwise the move is rejected. (This constraint is included because there will be moves which create stubborn coordination defects, and we found that these defects hindered the progress of MC annealing where the simulation would get ``stuck'' in local minima.) If the constraints are met, the new structure is accepted with the following probability:
\begin{equation}
    w=\begin{cases}
    1, & \text{if $\Delta E\le0$},\\
    \exp{(-\beta\Delta E)}, & \text{if $\Delta E>0$},
    \end{cases}
\end{equation}
where $\beta = (k_{\rm B}T)^{-1}$ and $\Delta E = E_{\text{new}} - E_{\text{old}}$. If $\Delta E>0$, we generate a random number $z$ in $[0,1)$, and if $w>z$, then the move is accepted, else it is rejected and the previous configuration is kept. In this work, $\beta$ has no physical relation to temperature since $\Delta E$ is not measured for all atoms about their equilibrium position, as the atoms are being held fixed after the bond transposition and thus thermal fluctuations are ignored (unlike with MD). Fixing particle positions at every move results in ergodicity being broken and thus samples are not taken from a Boltzmann distribution.\cite{Vink2014} Furthermore, the local-energy and NN-energy framework explicitly do not include all atomic energies which is another source of ergodicity being broken. This is not a concern for the present work as temperature-driven dynamics are not relevant (see Ref.\ \citenum{Vink2014} for an approach to introducing ergodicity to a similar problem) and results in $\beta$ becoming a tunable parameter. $\beta$ was initially chosen to be 2.0 eV$^{-1}$ to correspond to the study by Toh et al.,\cite{toh_synthesis_2020} and then tuned heuristically for the local- and total-energy-based MC runs. 

We use the atomic energies for the defect pair in the Metropolis criterion, defining the local defect energy as
\begin{equation}
    \varepsilon_{\text{local}} = \varepsilon_{1} + \varepsilon_{2}
\end{equation}
and the NN defect energy as
\begin{equation}
    \varepsilon_{\text{NN}} = \varepsilon_{\text{local}} + \sum_{i}^{\text{NN}}\varepsilon_{i},
    \label{equation:NN}
\end{equation}
where $\varepsilon_{i}$ is the local atomic energy of the $i$-th atom, with $i=1$ and $i=2$ denoting the two atoms in the SW defect pair, and $i=3..6$ the topological NNs of the SW defect pair (Fig.\ \ref{fig:cartoon}c--d). MC runs based on total energies were also carried out for reference. For the 200-atom systems, 25 independent and parallel simulations were conducted over 10,000 MC steps each for different $\beta$ values and for the 612-atom system, 20,000 MC steps were taken.

\begin{figure*}
\includegraphics[width=\textwidth]{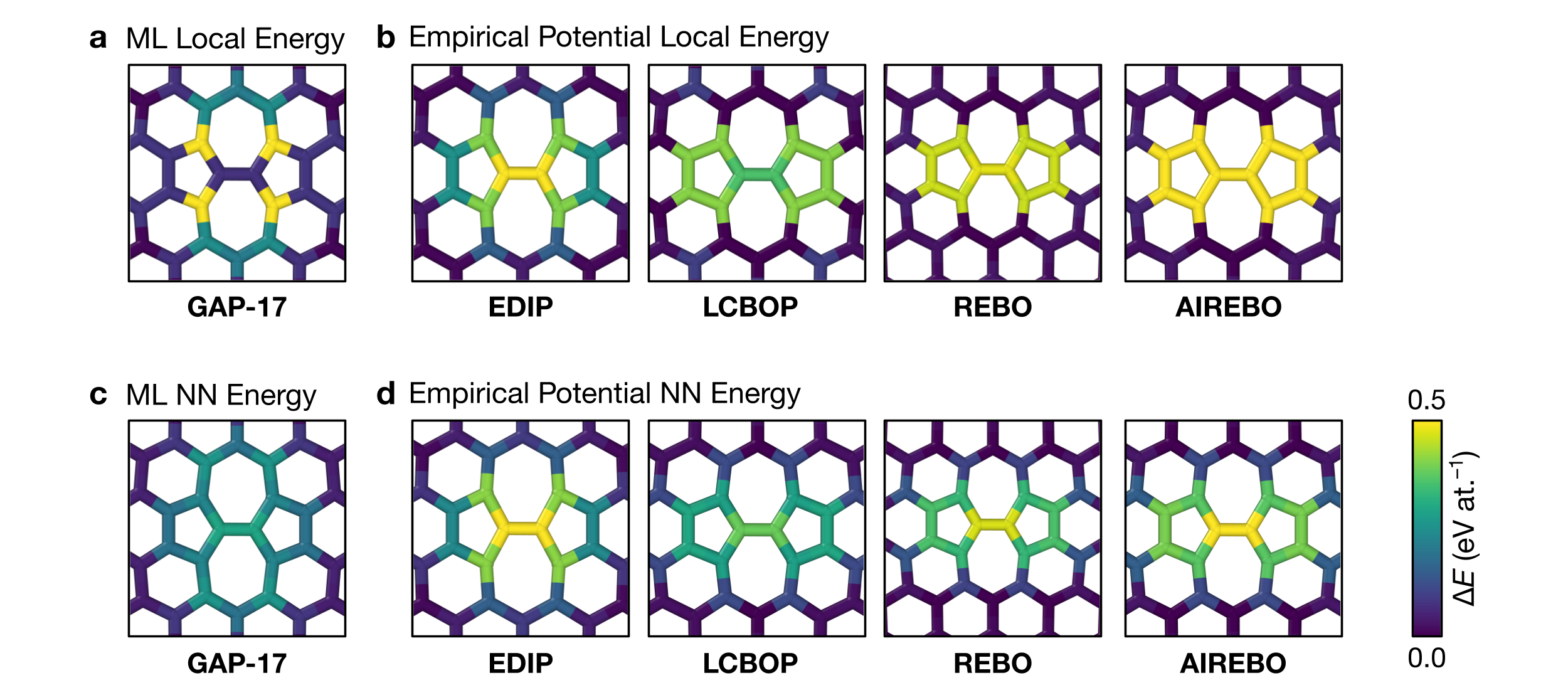}
\caption{
Structure and local energy of a Stone--Wales defect in graphene.
The figure compares local energies ({\em top row}) and nearest-neighbor averaged energies (``NN'', {\em bottom row}) for a single Stone--Wales (SW) defect as relaxed using the respective interatomic potential. (a--b) Local atomic energies from GAP-17 and empirical potentials, respectively. (c--d) Same but for NN energies (Eq.\ \ref{equation:NN}), from GAP-17 and empirical potentials, respectively. 
} {\label{fig:SW_all}}
\end{figure*}

\subsection*{\label{sec:Struct_analysis}Structural analysis} 

Ring statistics were determined using a shortest-path algorithm\cite{ring_count} as implemented in \texttt{Matscipy}.\cite{matscipy} Topological metrics typically used in network analysis were applied to the 612-atom structures, namely, Lemaître's law\cite{Lemaitre_1992} and the assortativity\cite{Newman2002} (as discussed below). For SOAP structural analysis, we compared each individual atom in a given aG structural model to an atom in cG. The SOAP parameters were: radial cutoff, $5.5 \angstrom$; cutoff transition width, $0.5 \angstrom$; neighbor-density smoothness, $\sigma_{\rm atom} = 0.5 \angstrom$; basis-set convergence parameters, $n_{\text{max}} = l_{\text{max}} = 16$; dot-product kernel raised to a power of $\zeta=4$. All structures were visualized using OVITO. \cite{Stukowski2010}

\section*{\label{sec:Results}Results and Discussion}

\subsection*{\label{sec:Local_e}Local energies for a single defect}

\begin{table}[t]
\caption{Structure and formation energy of a single SW defect in graphene from ML and empirical potentials. The table shows the computed bond length in pristine graphene ($d_{0}$), the defect-pair ($d_{1}$) and defect--NN bond lengths ($d_{2}$), the defect bond angle ($\theta$; cf.\ Fig.\ \ref{fig:cartoon}b), and three energies associated with defect formation: the local energy change in the rotated atoms only ($\Delta\varepsilon_{\text{local}}$), the NN energy change ($\Delta\varepsilon_{\text{NN}}$), and the total defect formation energy ($\Delta E_{\text{total}}$).}
\label{tab:Table1}
\begin{tabular}{@{}lccccccc@{}}
\hline
\hline
   & $d_{0}$ & $d_{1}$ & $d_{2}$ & $\theta$  & $\Delta\varepsilon_{\text{local}}$ & $\Delta\varepsilon_{\text{NN}}$ & $\Delta E_{\text{total}}$ \\
   & (\AA{}) & (\AA{}) & (\AA{}) & (deg) & (eV) & (eV) & (eV) \\
\hline 
GAP-17                       & 1.41 & 1.32 & 1.44 & 121.89 & 0.14                       & 2.25                    & 5.58                       \\
\hline
EDIP                        & 1.50 & 1.39 & 1.58 & 118.26 & 1.69                       & 3.32                    & 5.58                       \\
LCBOP                        & 1.42 & 1.34 & 1.45 & 121.76 & 0.73                       & 2.39                    & 5.33                       \\
REBO                         & 1.42 & 1.34 & 1.46 & 120.64 & 0.99                       & 2.85                    & 5.61                       \\
AIREBO                       & 1.40 & 1.32 & 1.44 & 120.88 & 1.09                       & 3.15                    & 6.27                       \\
\hline 
\hline
\end{tabular}
\end{table}

Figure \ref{fig:SW_all} characterizes the structure and energetics of a single SW defect in graphene. GAP-17 predicts that the energies of the SW defect pair of atoms are considerably {\em lower} in energy than those of the NN atoms. This behavior is in marked contrast with that for the empirical potentials (Fig.\ \ref{fig:SW_all}b and Table \ref{tab:Table1}): the latter suggest that the SW-pair atoms are relatively {\em higher} in energy (with the exception of LCBOP), with atoms in the 5-membered rings being higher in energy overall and those in 7-membered rings being lower. LCBOP and GAP-17 predictions are similar when averaged over the pentagon-pair environments, only differing slightly in structural parameters. REBO and AIREBO show slightly larger SW-pair energies and more strained structures (Table \ref{tab:Table1}).

\begin{figure*}
\includegraphics[width=0.9\textwidth]{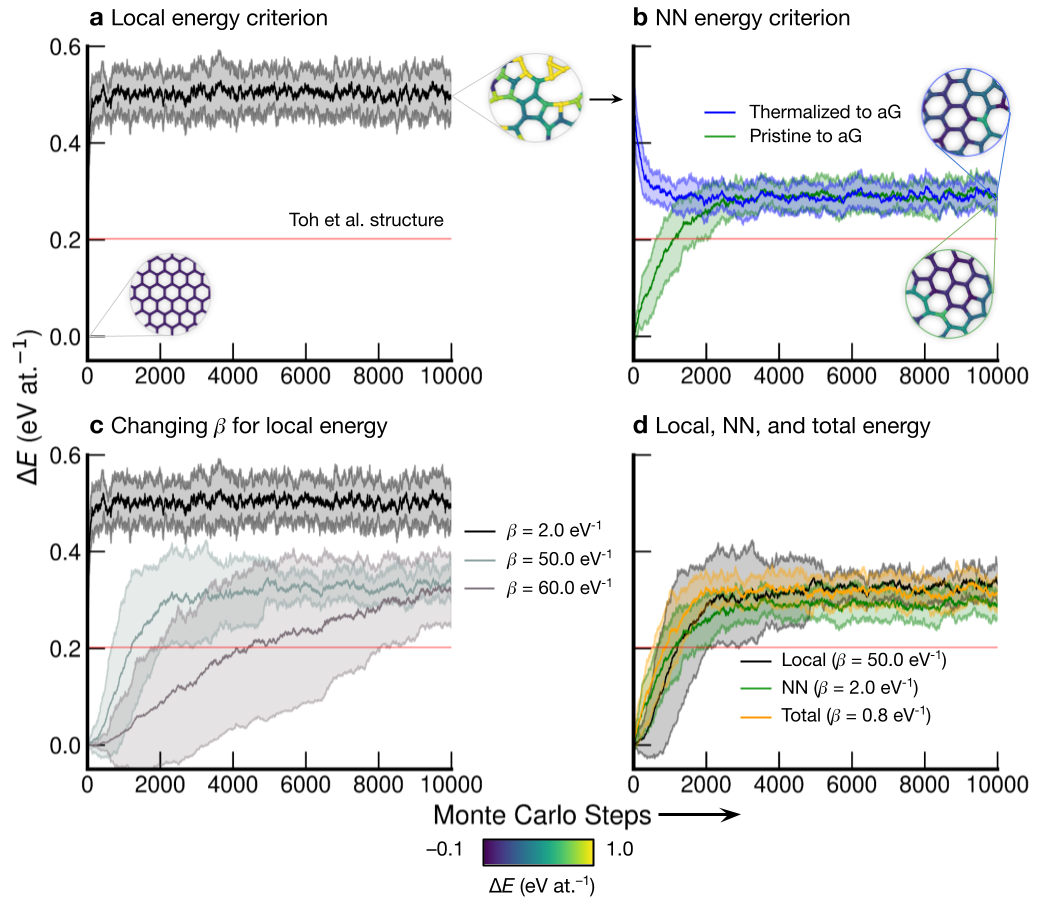}
\caption{
Evolution of disordered graphene structures during Monte-Carlo simulations.
Lines indicate mean energies for an ensemble of 25 separate runs, with shaded regions indicating a standard deviation of 1$\sigma$. Representative structural snapshots are shown, with atoms color-coded by local energy. (a) The atomic energies of the two atoms involved in the SW transformation are used for the Metropolis criterion, creating a ``thermalized'' structure rapidly at $\beta = 2.0$ eV$^{-1}$. These structures are used as initial configurations in (b). (b) MC annealing via NN energies ($\beta = 2.0$ eV$^{-1}$) for the Metropolis criterion. The curves converge at slightly above the energy of an MC-annealed paracrystalline structure from Toh et al.\ (Ref.\ \citenum{toh_synthesis_2020}), which is included for reference (red line). (c) The effect of $\beta$ on the MC energy profile. It is seen that a smaller $\beta$ (higher simulation ``temperature'') leads to a higher-energy structure and faster convergence. (d) Heuristically tuning $\beta$ to find a mutual limiting distribution between MC approaches (within statistical fluctuations).}{\label{fig:conv}}
\end{figure*}

The total defect formation energy $\Delta E_{\text{total}}$ for GAP-17 is similar overall to that predicted by most empirical potentials. EDIP, LCBOP, and REBO were parameterized for $\text{sp}^{2}$ environments, and their $\Delta E_{\text{total}}$ is in close agreement with that for GAP-17. AIREBO was also parameterized for $\text{sp}^{2}$ environments, however the added long-range term may not be optimized for 2D planar carbon. In contrast, $\Delta\varepsilon_{\text{local}}$ and $\Delta\varepsilon_{\text{NN}}$ are not in close agreement, where $\Delta\varepsilon_{\text{local}}$ differs the most across all the potentials (except for REBO and AIREBO which are parameterized identically for short-range terms). Overall, GAP-17 predicts the lowest energy for the NN defect formation energy among the potential models investigated.

Structural parameters as predicted by the respective potentials are also shown in Table \ref{tab:Table1}. The defect-pair bond length ($d_{1}$), defect--NN bond length ($d_{2}$), and defect bond angle, $\theta$ (Fig.\ \ref{fig:cartoon}b) are in close agreement across all potentials---with the exception of EDIP, which notably increases the cell parameters upon relaxation. This is shown by the difference in cG bond length ($d_{0}$) between EDIP and the other potentials when the cell parameters are allowed to relax, with EDIP predicting an elongated $d_{0}$. GAP-17 and LCBOP are in close agreement with regard to $\theta$, predicting a slightly wider bond angle for the $5\vert7$ pair, whereas EDIP predicts it slightly lower, notwithstanding the fact that the EDIP defect formation energy is in close agreement with that for GAP-17. 

\subsection*{\label{sec:MC_anneal}Using local energies to drive Monte-Carlo annealing}

Figure \ref{fig:conv} shows the energy profiles of independent parallel MC simulations via successive SW transformations, which we use to create structural models of aG. Panel (a) shows the evolution of the average energy of the ensemble when using (only) the atomic energies of the SW defect pair in the Metropolis criterion. Clearly, using a $\beta$ value of 2.0 eV$^{-1}$ leads to a highly disordered 3-coordinate structure (as shown in the inset), with energy convergence occurring after about 1,000 MC steps. Such structures contain high-energy 3- and 4-membered rings with significant strain. We call the resulting structures ``thermalized'' and use them as starting points for the runs characterized in blue in Fig.\ \ref{fig:conv}b, as indicated by an arrow. 

Nearest-neighbor (NN) atoms are defined topologically in the present work, and so we include the sum over the SW pair and its bonded neighbors, including six atoms in total (Eq. \ref{equation:NN}). The difference before and after the SW transformation is used in the Metropolis criterion. Results from two protocols, starting either from cG or from the previously thermalized structures, converge to similar energy values (green and blue curves in Fig.\ \ref{fig:conv}b). It seems that runs from both starting points tend toward a mutual limiting distribution. This is reflected in Fig.\ \ref{fig:conv}b, as both curves have means and standard deviations in very close agreement after approximately 4,000 MC steps. However, it is apparent that this protocol is not ergodic, as discussed in the Methods section. The principle of ergodicity states that all possible configurations of the system should be attainable\cite{landau2005guide} and thus it is clear that given the topological constraints imposed, it is impossible for certain proposed moves to be accepted even though they may be energetically favorable. 

Figure \ref{fig:conv}c illustrates the effect of varying $\beta$ on the resulting energy profile when using the local-energy criterion. It is clear that using a lower $\beta$ value corresponds to a higher-energy final structure as well as rapid convergence: for $\beta = 2.0$ eV$^{-1}$, convergence occurs in fewer than 1,000 MC steps (black line). Increasing $\beta$ results in lower-energy structures and slower convergence, as shown by the $\beta = 50.0$ eV$^{-1}$ and $\beta = 60.0$ eV$^{-1}$ results. For $\beta = 60.0$ eV$^{-1}$, convergence has not occurred after 10,000 steps, however the mean appears to be tending towards that for $\beta = 50.0$ eV$^{-1}$, implying that there might be a minimum level of disorder that is ``stable'' for the local-energy criterion. Additionally, as $\beta$ was increased beyond $60.0$ eV$^{-1}$, proposed moves were always rejected, further suggesting a minimum convergence energy using this framework. 

Figure \ref{fig:conv}d shows the results from three different runs with heuristically tuned $\beta$ values for local-, NN-, and total-energy criteria, respectively, converging towards a mutual distribution (within statistical fluctuations). This result allows the direct comparison of structures generated by the different frameworks side-by-side, as the effect of $\beta$ has been removed. Such a comparison is particularly instructive for lower-energy structures, as we will show below: they show a richer configurational space, including medium-range order (as compared to the highly disordered high-energy structures), and this space may be traversed differently by different MC runs.

\subsection*{\label{sec:level2}The configurational space of amorphous graphene}

\begin{figure*}
\includegraphics[width=0.9\textwidth]{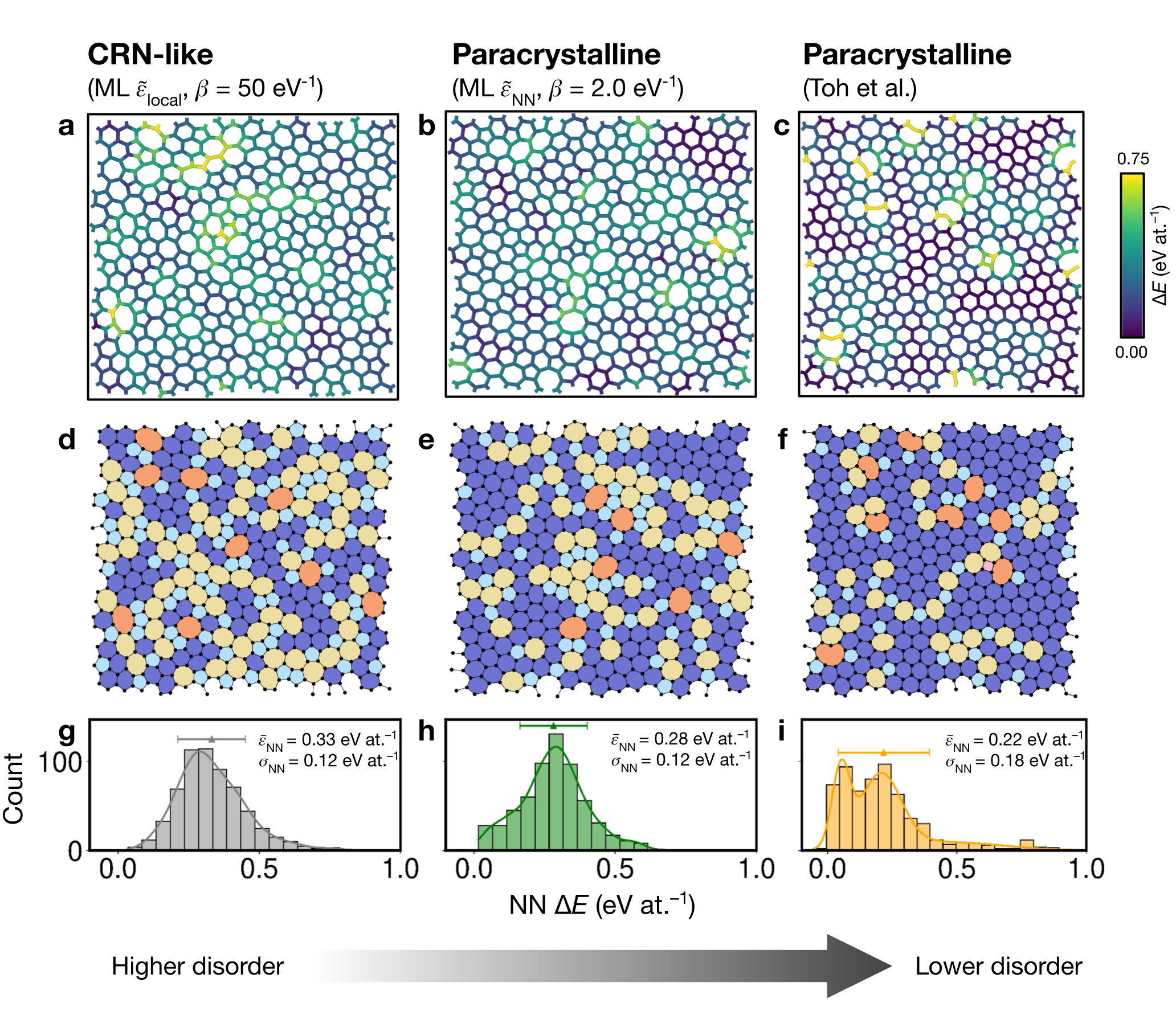}
\caption{Structural models of amorphous graphene. Structures in panels (a--c) are color-coded according to the average atomic energy over a chosen atom and the corresponding NNs. (a) Structure after the final MC step for the local-energy criterion and $\beta=50$ eV$^{-1}$.  (b) As for (a), but with NN energies used in the Metropolis criterion. (c) Structure taken from Toh et al.\cite{toh_synthesis_2020} and optimized with GAP-17. (d--f) Structures as in (a--c), now color-coded by ring size (pink, $\le 4$; light blue, 5; dark blue, 6; yellow, 7; orange, $\ge 8$). (g--i) Distributions of NN energies in the respective structures, with mean values and standard deviations given. Curves were obtained as kernel density estimates.
}{\label{fig:struct}}
\end{figure*}

Having explored different protocols for generating aG structural models, we next created larger-scale structures: the system size was increased to 612 atoms and structures from local- and NN-energy-based searches were studied. Additionally, the 610-atom structural model published by Toh et al.\ (Ref.\ \citenum{toh_synthesis_2020}) was included for comparison. Figure \ref{fig:struct} shows these three configurations. 

For the local-energy framework, the absence of crystal-like pockets is clear as seen in panels (a) and (d). The structure resembles a CRN, with chains of 5- and 7-membered rings running across the cell, and there are more large rings compared with the other structures. The pronounced disorder is likely a result of the low ML energy for the SW defect itself (Fig.\ \ref{fig:SW_all}a), and thus of the low energy cost for these transpositions (given that medium-range order is not captured in the local energies alone). 

With NN energies used in the Metropolis criterion, panels (b) and (e) suggest that using an NN criterion encourages small pockets of crystal-like regions forming, indicating that direct contributions from NNs maintain medium-range order, whilst retaining an amorphous structure at $\beta=2.0$ eV$^{-1}$. Visually, the locally averaged energies (up to NNs) in panels (a--c) suggest that using just the local defect-pair energy gives more regions of higher energy (shown in yellow in Fig.\ \ref{fig:struct}a) compared with the NN framework (Fig.\ \ref{fig:struct}b). This appears consistent with the nature of CRNs versus paracrystalline structures, and color-coding by NN energies shows a clear difference between the two structures.

In the NN-energy-based structure (Fig.\ \ref{fig:struct}b), there are pockets of crystallinity, indicated by regions of ordered 6-membered rings. Interestingly, there is an aggregation of more disordered regions, with 5-membered rings surrounding larger 7- and 8-membered rings. For the structure shown in Fig.\ \ref{fig:struct}c, the authors started from a randomized, hard-sphere constrained structure and worked towards a paracrystalline sample using the AIREBO potential. In the resulting structure, we note the presence of coordination defects, since the randomized initial structure was not topologically constrained. There is also a 4-membered ring (pink in Fig.\ \ref{fig:struct}f). It is evident that this sample is paracrystalline with regions of locally crystal-like order separated by $5\vert7$ grain boundaries, and that it is more ordered than the NN-based structure (Fig.\ \ref{fig:struct}e). Larger defects appear to congregate as seen, for example, in the top of the figure. These defective environments are high in energy, as shown by the color-coding.

Figure \ref{fig:struct}g--i shows the distribution of NN energies for the respective structures. The average energy for the atoms in the local-energy-based structure is 0.33 eV at.$^{-1}$ with a standard deviation of $\sigma=0.12$ eV at.$^{-1}$, and the energy of the NN-based structure is $0.28 \pm 0.12$ eV at.$^{-1}$ above pristine graphene. Both values are close to the median energy resulting from the independent runs for the 200-atom system (Figs.\ \ref{fig:conv}b and \ref{fig:conv}d): hence, an individual 200-atom run will not likely be sufficient to describe aG, but an ensemble of multiple independent runs will be---just like many small-scale structural models of 3D amorphous carbon have been used in the fitting of GAP-17. \cite{GAP17} The distribution in the locally averaged energies is similar between Figs.\ \ref{fig:struct}g and \ref{fig:struct}h.

\begin{figure*}
\centering
\includegraphics[width=0.9\textwidth]{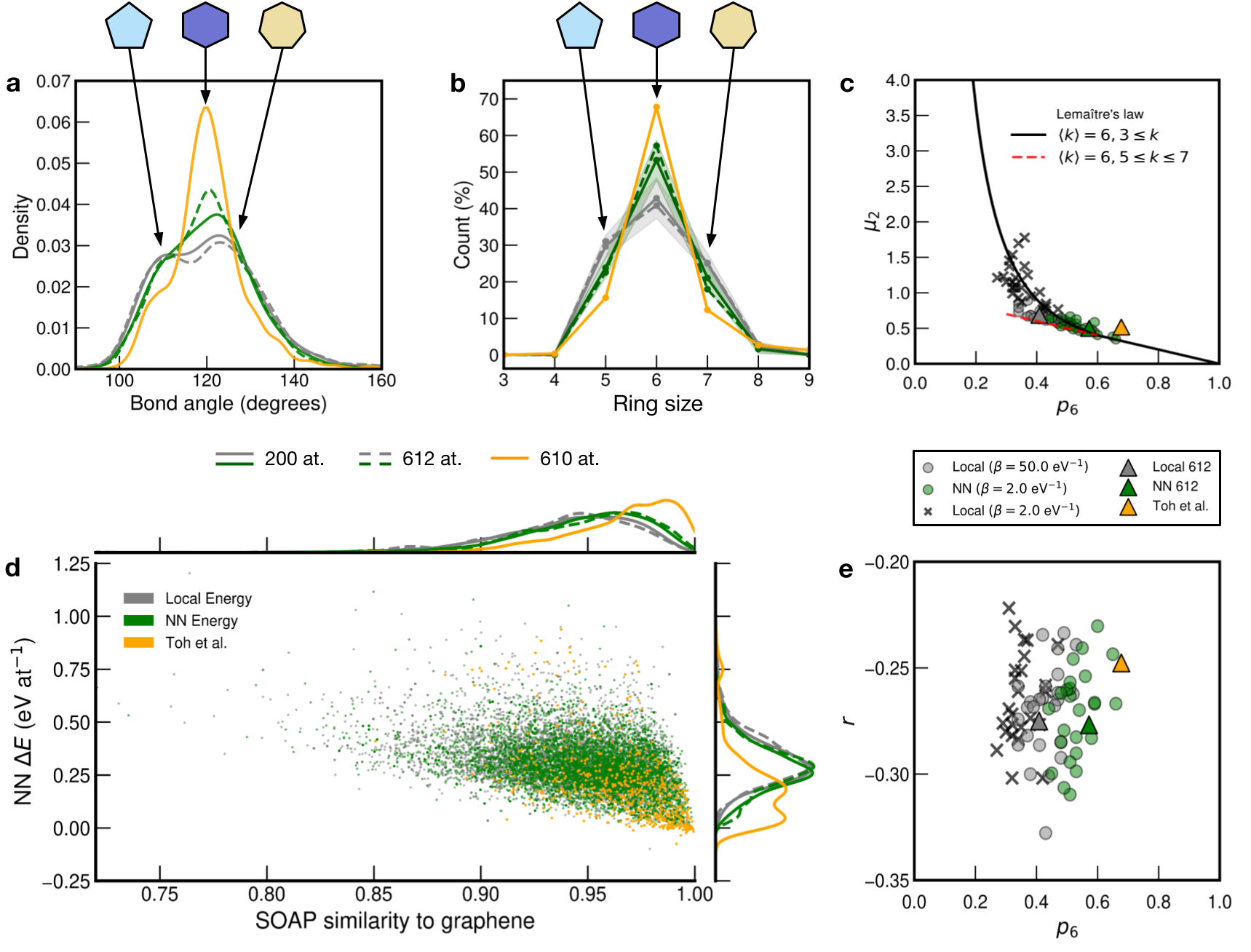}
\caption{
Local structure and stability in models of amorphous graphene. Results for 25 local-energy-based ($\beta=50.0$ eV$^{-1}$, grey) and NN-energy-based ($\beta=2.0$ eV$^{-1}$, green) runs, i.e., for $25 \times 200$ atoms each, and for the 612-atom structures based on local ($\beta=50.0$ eV$^{-1}$, gray dashed) and NN ($\beta=2.0$ eV$^{-1}$, green dashed) energies are given. The structural model from Toh et al.\ is also analyzed (orange, 610 atoms). Data from the 612-atom structures and Toh et al.\ are scaled arbitrarily to fit. (a) Bond-angle distributions for local, NN and Toh et al. (b) Ring statistics for the different structures. 
(d) An NN energy vs SOAP plot, following Ref.\ \citenum{Si_local}.
Kernel density estimates (KDEs) are used to show distributions of properties. The bandwidth is determined following Scott's rule, \cite{scott2015multivariate} with a grid size of 200. Additionally, panels (c) and (e) show plots of network-topology metrics, namely, Lemaître's law and assortativity, respectively. 
}{\label{fig:e_soap}}
\end{figure*}

For Fig.\ \ref{fig:struct}i, we find a bimodal distribution in the energy, indicating that this structure shows higher paracrystalline order compared to that generated by the NN framework. There exist pockets of more ordered, more energetically favorable regions in between more disordered ones, as seen in the structural model shown in panels (c) and (f). Very recently, a paracrystalline sample of diamond has been synthesized.\cite{Tang2021} This discovery, along with the synthesis and characterization of paracrystalline graphene \cite{toh_synthesis_2020, Barkema2022} suggests that the landscape of disorder in carbon may have a link between ``fully'' amorphous and crystalline. This is further reflected in the NN energy distributions in Fig.\ \ref{fig:struct}, where panel (g) clearly shows a single distribution, panel (i) a bimodal distribution, and panel (h) a small hint of one. These ML energy distributions may provide a quantitative distinction between CRN and paracrystalline graphene, where a prominent bimodal distribution indicates the latter and the lack thereof indicates the former. 

We show further, quantitative structural indicators in Fig.\ \ref{fig:e_soap}. The bond-angle distribution (panel a) characterizes medium-range order as there will be characteristic peaks if there is significant ordering. The local-energy-based curves (gray) display a wide range of bond angles, reflecting the relatively high level of structural disorder. The absence of a clear peak at $120 \degree$ also emphasizes the reduction of medium-range order as is characteristic of CRNs. The NN curves (green) are narrower and centered around $120 \degree$, consistent with locally crystal-like environments. A shoulder peak at $\approx 109 \degree$ shows the strong presence of 5-membered rings. The Toh et al. structure has a pronounced peak at $120 \degree$ with a smaller shoulder at $\approx 109 \degree$, indicative of the larger degree of locally crystal-like order. The bond angles in the 200-atom models (solid lines) agree well with those in the 612-atom ones (dashed).

The count of shortest-path rings is another metric for medium-range order. Ring counts for the local-energy MC runs reflect the large disorder, showing more 5- and 7-membered rings than in the other structures. The structures from the NN runs have ring counts centered around 6, as for the Toh et al.\ structure, with a larger count suggesting increased crystallinity in the latter case. As with the bond-angle distributions (Fig.\ \ref{fig:e_soap}a), the ring statistics for the 200-atom versus 612-atom structures are in close agreement within each other, within the standard deviation of the values for the former (Fig.\ \ref{fig:e_soap}a).

SOAP is a structural similarity metric for atomic environments. \cite{SOAP2013} 2D plots can reveal correlations between SOAP similarity on the one hand, and locally averaged energies on the other hand.\cite{Si_local} Figure \ref{fig:e_soap}d shows the distribution of locally averaged energies (up to NN) and SOAP similarity to cG. The wider the spread in both axes, the greater the structural disorder. Comparing local-energy- with NN-energy-based data, both KDE curves are in close agreement, being skewed slightly toward the left (SOAP) and to higher energies for the runs using local energies only. This is expected, since the ML model predicts relatively low energies for the SW pair (Fig.\ \ref{fig:SW_all}), allowing structures to become more disordered. When averaging over NN environments, the distribution shifts and narrows as we observe fewer highly disordered environments. Data for the structure from Toh et al.\cite{toh_synthesis_2020} are provided to show how ordered this structure is. Defective environments are clearly identified with a series of data points at higher energies. As with Fig.\ \ref{fig:struct}, there is a clear bimodal distribution in the KDE curve. As seen for the structural indicators characterized in Fig.\ \ref{fig:e_soap}a--b, the dashed lines representing the 612-atom structures agree well with the set of separate 200-atom structures. 

In addition to the well-established ring statistics and the SOAP similarity, we analyze the structural models with two topological metrics typically used in network theory, viz.\ Lemaître's law\cite{Lemaitre_1992} and the assortativity metric.\cite{Newman2002} Lemaître's law connects the second moment of the ring distribution ($\mu_{2}=\langle k^{2} \rangle - \langle k \rangle^{2}$, where $k$ is the ring size) with the fraction of six-membered rings, $p_{6}$. Since we impose the constraint that all atoms must be 3-fold connected, it follows that the distribution can be explained well by a single maximum-entropy distribution leading to a characteristic curve.\cite{Lemaitre_1992} As $\lim_{p_{6} \to 0.6}$ from $p_{6}=1$, $\mu_{2}$ increases linearly as $1-p_{6}$ in the region of $p_{6}\gtrapprox 0.6$. This line is extended (red line) to show that most of the points from both local ($\beta=50.0$ eV$^{-1}$) and NN frameworks are located on this curve. This curve corresponds to the maximum-entropy solution if only 5-, 6-, and 7-membered rings are present, which is the case in both these frameworks. $\mu_{2}$ then increases exponentially beyond this where structures from the local framework at $\beta=2.0$ eV$^{-1}$ are located. These structures show greater disorder and a wider spread of $\mu_{2}$ values, as expected due to the presence of 3- and 4-membered rings, where the maximum-entropy solution follows an exponential profile. For the 612-atom structures, the local framework ($\beta=50.0$ eV$^{-1}$) based structure is more disordered than the NN-based one and the Toh et al.\ structure (orange). The latter does not yield a datapoint on the curve as it has a (small) number of 2-coordinate sites and hence would be located on a different Lemaître curve.\cite{ormrod_morley_controlling_2018} 

The assortativity, $r$, measures how likely a large ring is to be next to a smaller ring (disassortative, $r < 0$) or to other large rings (assortative, $r > 0$). In many physical systems, one finds a preference for disassortative configurations. This is, here, reflected in the range of $r$ values (Fig.\ \ref{fig:e_soap}e). For the 200-atom systems, we find no discernible correlation between $r$ and $p_{6}$. For the 612-atom ones, both the local and NN frameworks yield structures with similar $r$ values. The Toh et al.\ structure has a slightly less negative $r$, implying a slight preference for a more random arrangement. This may be due to coordination defects in the structure skewing $r$ toward zero.

\section*{Conclusions}

Atomic energies predicted by the GAP ML framework can be used to drive Monte-Carlo structural exploration in principle. We have shown this by creating structural models of amorphous graphene, one of the prototypical disordered systems in physics and chemistry. We found that using (only) ML atomic energies leads to structures resembling continuous random networks, whereas including nearest-neighbor energies tends to drive the simulations toward some degree of paracrystalline order. We suggest that histograms of local energies, as shown in Fig.\ \ref{fig:struct}, can give insight into the degree of ``randomness'' in different amorphous networks: deviations from random order are visible as peaks at either low (paracrystalline) or high (coordination defects) energy.

The demonstrated ability to use ML local energies in MC annealing indicates potential for future research on amorphous materials. Describing local environments using ML methods can provide insight into the relation between atomic structure and energetics, and therefore structural stability, in amorphous materials---including amorphous carbon, which has emerging applications in biosensing \cite{Laurila2017} or batteries. \cite{Olsson2022} The fact that NN-averaged energies yield reasonable, partly paracrystalline structural models may be attributed to the fact that they provide ``smoothing'' over the variance in local atomic energies. This finding is consistent with earlier findings for the electronic DOS \cite{BenMahmoud2020, deringer_origins_2021} and might have wider consequences for ML predictions of local properties, which are yet to be fully explored.

\begin{acknowledgments}
We thank Prof.\ G.\ Cs\'anyi for helpful comments on the Monte-Carlo simulations. We are grateful for support from the EPSRC Centre for Doctoral Training in Theory and Modelling in Chemical Sciences (TMCS), under grant EP/L015722/1. V.L.D. acknowledges support from the Engineering and Physical Sciences Research Council through a New Investigator Award [grant number EP/V049178/1]. This paper conforms to the RCUK data management requirements. The authors would like to acknowledge the use of the University of Oxford Advanced Research Computing (ARC) facility in carrying out this work (http://dx.doi.org/10.5281/zenodo.22558).
\end{acknowledgments}

\section*{Data availability}

Data supporting this work will be provided openly upon journal publication.

\section*{References}
\providecommand{\noopsort}[1]{}\providecommand{\singleletter}[1]{#1}%

\end{document}